
 \def\R{{\rm I\mathchoice{\kern-0.70mm}{\kern-0.70mm}{\kern-0.65mm}%
  {\kern-0.50mm}R}}
\magnification = \magstep1
\pageno=0
\rightline{UMD preprint \# 92-176}
\rightline{submitted to Phys. Rev. D}
\null \vskip 3 pt plus 0.2 fill
{\centerline{\bf Splitting of an Extremal Reissner-Nordstr\"om Throat}}
\medskip
{\centerline{\bf via Quantum Tunneling}}
\vskip 3 pt plus 0.2 fill
{\centerline{Dieter Brill}}
\vskip 3 pt plus 0.1 fill
\it{\centerline{Department of Physics}}
{\centerline{University of Maryland}
{\centerline{College Park, MD 20742}}
{\centerline{Bitnet: Brill@UMDHEP}}
\vskip 3 pt plus 0.3 fill
\rm
{\centerline{\bf {Abstract}}
\medskip
The interior near the horizon of an extremal Reissner-Nordstr\"om
black hole is taken as an initial state for quantum mechanical
tunneling. An instanton is presented that connects this
state with a final state describing the presence of several horizons.
This is interpreted as a WKB description of fluctuations due to the
throat splitting into several components.
\smallskip
\rightline{\it ``Fluctuat nec mergitur"}

\vfill \eject
\vskip 0.5truein
{\centerline{I. INTRODUCTION}}
\vskip 0.25truein
The significance of geometries of non-Euclidean topology in General
Relativity was first spelled out in Wheeler's
``geometrodynamics" [1].  Localized topological complications,
such as  Wheeler
wormholes, are allowed even at the classical level.
However, the  Einstein dynamics of classical gravity coupled to
classical matter fields provides no compelling reason that such
non-Euclidean topologies should occur in nature, because the
topology of spacelike Cauchy surfaces is conserved in the classical
Einstein time development [2].  Wheeler has argued that
in quantum geometry,  wormholes can be created and destroyed,
for example as part of the processes that constitute the vacuum
fluctuations. According to this view the creation or destruction
of a wormhole can be regarded as
a quantum tunneling through a classically forbidden region.

Tunneling processes can be treated as propagation in imaginary
(``Euclidean") time [3], and this description is appropriate for
quantum fluctuations of topology
not least because the kinematical description of topological changes
is simple in Riemannian\footnote{$^1$}{In this paper we reserve the word
Euclidean (without quotes) for $\R^n$ with the flat geometry
whose $n=2$ or $3$ version Euclid would recognize as the subject of his
``$\Sigma \tau o \iota \chi \varepsilon \tilde \iota \alpha"$.
More general manifolds with a positive definite metric
will be called Riemannian, as opposed to spacetimes with an indefinite
metric, which will be called Lorentzian.} manifolds.
For example, for any pair of three-dimensional topologies one can find
a Riemannian 4-manifold that interpolates between them.
This is true for Lorentz manifolds only if one accepts closed timelike
curves [4].
Another, practical reason for using the Euclidean description as
part of a stationary phase (WKB) approximation is that
we can have at least some confidence in it, lacking as we do today
a complete theory of quantum gravity.

To implement this approximation one starts by looking for
Riemannian  solutions of the gravitational (and matter)
field equations (``instantons"). An instanton can be considered as
a tunneling history in the same way that a Lorentzian solution of
the field equations represents an actual (classical) history of the
fields. To make the development of this history more explicit one
usually ``slices" a space-time by spacelike surfaces, and one can
similarly slice a 4-dimensional instanton by non-intersecting
three-dimensional surfaces (some of which may be singular). The
interpretation of the instanton as an approximate description
of a wave functional
in the 3-metric representation of quantum geometry is also analogous
to that of classical (Lorentzian) solutions: In situations described
by the latter, the wave functional $\Psi [^3{\cal G}]$ is sharply
peaked about Riemannian 3-geometries $^3{\cal G}$ that all fit (as
spacelike surfaces) into one {\sl Lorentzian\/} 4-geometry [5];
in the  instanton case, $\Psi[^3{\cal G}]$ is sharply
peaked about 3-geometries that all fit into one {\sl Riemannian\/}
4-geometry [6]. The geometries that are connected by such a tunneling
history are contained in the instanton 4-geometry as follows:
the asymptotic region corresponds to the initial metastable
state, and a maximally geodesic 3-surface,
which has zero geometrical momentum as measured by the
extrinsic curvature tensor, corresponds to the final
state ---  the moment at which the decay product emerges.
(Because the momenta vanish, the latter state is  a suitable
initial state for a Riemannian as well as for a Lorentzian
development, and corresponds to the classical turning point or
``bounce" in potential motion.)  This interpretation has an obvious
generalization to the case when matter fields are also present.

A number of instanton solutions have been interpreted in this
way, for example the production of Wheeler wormholes by a magnetic
field [7]. However, to describe fluctuations of a wormhole that
is already present one would expect an instanton related in some limit to
a classical black hole background, for example the Schwarzschild-Kruskal
solution. The very presence of a horizon, which characterizes
black holes, usually prevents the existence of a suitable instanton,
because in the analytic continuation to Euclidean time the manifold
is regular at the horizon only if this time is a periodic coordinate
[8]. Thus the manifold cannot have the desired asymptotic behavior
in imaginary time. Equivalently, the problem is that a classical
wormhole is not a stationary state, since its throat collapses,
whereas the simple instanton picture of a classically forbidden
decay is applicable only if the
initial and/or final states are stationary. (The periodic
instantons connected with black holes do have a physical interpretation,
but it has to do with thermal phenomena such as Hawking radiation.)

There is however one type of black hole that does not offer these
problems, namely the ``extremal Reissner-Nordstr\"om" black hole
and its generalizations. Here the ususal two Killing horizons of charged
black holes coincide (degeneracy in the sense
of Carter [9]).  Two consequences are that the horizon is
located at infinite distance on the spacelike surface $t =$ const, and
that, in the analytic continuation, the Euclidean time  does not
have a finite period. Thus the existence of asymptotic
regions of differing topology becomes possible.\footnote{$^2$}{In the
thermal interpretation the infinite period means that the temperature
vanishes. Thus a scenario is possible in which all black holes
that have magnetic charge and lose mass by
Hawking radiation become magnetically charged
extremal black holes. The fluctuations discussed here would then
take over near this typical endpoint of the thermal evolution.}

The analogous picture
in potential motion is given by a potential that has two degenerate minima
at $x_1, x_2$. Here the motion in Euclidean time
starts with an asymptotically slow roll-off from $x_1$, and ends with an
asymptotically slow ascent to $x_2$ --- there is no bounce back to $x_1$.
In this case the instanton does not signal an instability where
after a sufficiently long time the particle is found in the
initial state with vanishingly small probability. Instead, the
particle is found with some probability in either of the degenerate
minima. The action of the instanton is related in the WKB approximation
to the energy splitting between the two classically degenerate
ground states, and therefore to the
frequency with which the quantum system fluctuates between $x_1$ and
$x_2$ [10].

The present paper presents an instanton with the corresponding properties
in a spacetime theory. It can be considered to have two asymptotic regions
that can be foliated by 3-surfaces having topologies that differ between
the two regions.  Furthermore, in these asymptotic regions
the geometry of the 3-surfaces agrees with the asymptotic interior
of one respectively two (or several) extremal Reissner-Nordstr\"om
wormholes.\footnote{$^3$}{We use this term rather than ``black hole"
because we contemplate an interior free of any collapsing matter.}
The instanton may therefore be considered a description
of the topological fluctuations near the horizon of such
extremal Reissner-Nordstr\"om wormholes.

Section II recalls the asymptotic form of the metric and fields
near the extremal Reissner-Nordstr\"om
horizon (usually called the Bertotti-Robinson universe,
and abbreviated here as BR) and its
Euclidean version.  Section III gives the instanton solution that
connects one BR universe with several, and discusses
the quantities that are conserved during the transition.
In Section IV it is shown that the action of the instanton is
finite, and remarks are made about the probable form of an instanton
that would connect different numbers of wormholes not only in the
extreme interior, but everywhere.  Section V summarizes the conclusions.

\vskip 0.5truein
{\centerline{II. LIMITING FORM OF EXTREMAL REISSNER-NORDSTR\"OM GEOMETRY}}
\vskip 0.25truein
\nobreak
The well-known Reissner-Nordstr\"om metric for a magnetically
charged black hole, in the case $Q^2 = M^2$, takes the form,
in the usual static coordinates where $r \geq M$ [9]
$$ds^2_{RN} = (1 - M/r)^{-2} dr^2 + r^2 d\Omega ^2 - (1 - M/r)^2 dt'^2
\eqno(1a)$$
$$F_{RN} = M\, d(\cos\theta) \wedge d\phi. \eqno(1b)$$
The geometry on a totally geodesic spacelike surface $t' =$ const.~is
complete --- on such a surface the horizon $r = M$
is at infinite spatial distance
from any point with $ r>M$. Thus this surface does not have the
usual wormhole shape with two asymptotically flat regions; instead
it is asymptotically flat at $r \rightarrow  \infty$, but becomes an
infinitely long cylinder with a constant magnetic field along the
cylinder's generators in the limit $r \rightarrow  M$. In this limit it
coincides with the BR solution [11] that describes
this exact cylinder (infinite in both $+z$ and $-z$ directions),
$$ds^2_{BR} = M^2(dz^2 + d\Omega ^2 -\cosh ^2 z \ d{\tau '} ^2)
 \eqno(2a)$$

$$F_{BR} = M\, d(\cos \theta)\wedge d\phi. \eqno(2b)$$
(Whereas metric (2a) is  geodesically complete, neither (1a) nor
the Lorentzian counterpart of (5) below
are geodesically complete, because timelike or null geodesics can
reach the horizon at $r = M$ resp.~$\rho = 0$
in a finite affine parameter interval\footnote{$^4$}{for
a discussion of the analytic extension
and of the relation between these metrics and (2a), which is complete,
see Carter [9]}; but the Riemannian versions, (3a)
--- a special case of (8a) --- and (5)
will turn out to be geodesically complete.) The single parameter $M$
in (2) no longer has the interpretation as a mass, but it measures both the
(transverse) size of the BR universe and the (constant) field strength.
The total magnetic flux through the universe, $\oint F$, is $4\pi M$.
We will therefore call $M$ the flux parameter.

Since these metrics are static they can easily be continued to ``Euclidean"
time by defining $t' = it$ resp.~$\tau ' = i\tau$, which merely changes
the sign of the metrics' last terms:
$$ds^2_{RN} = (1 - M/r)^{-2} dr^2 + r^2 d\Omega ^2 + (1 - M/r)^2 dt'^2
\eqno(3a)$$
$$ds^2_{BR} = M^2(dz^2 + d\Omega ^2 +\cosh ^2 z \ d{\tau} ^2) \eqno(3b)$$
$$F_{RN} = F_{BR} = M\, d(\cos \theta)\wedge d\phi \eqno(3c)$$
In metric (3b) we now
introduce new coordinates [12]
$$\rho = Me^{-\tau}\cosh z, \quad t = Me^{\tau} \tanh z \eqno(4)$$
and find
$$ds^2_{BR} = (M/\rho)^2 d\rho^2 + M^2 d\Omega ^2 + (\rho/M)^2 dt^2.
 \eqno(5)$$
This is identical to (3a) in the limit $\rho = r-M \rightarrow 0$.

Equation (3b) shows that the Riemannian BR geometry is the direct product
of two spaces of constant curvature, the 2-sphere $S^2(\theta,\ \phi)$ and
the negative constant curvature hyperbolic space $H^2(z,\ \tau)$. The
coordinates are analogous in the sense that $\theta - {\pi \over 2}
\rightarrow iz$,
$\phi \rightarrow i\tau$ transform the two-dimensional metrics into
each other. The $\rho$, $t$ coordinates, on the other hand, are a type
of polar coordinates, with $t =$ const.~describing radial geodesics
originating from a point at infinity.
\vskip 0.5truein
{\centerline{III. AN EINSTEIN-MAXWELL INSTANTON}}
\vskip 0.25truein
\nobreak
To find instantons that describe the change of one wormhole, or
one BR universe, into
several would a priori seem difficult because the instantons would
be expected to have none of the continuous symmetries (except, in the
case of the BR universe, the $z$-translation) that ususally make
solutions of the Einstein or Einstein-Maxwell equations possible.
Remarkably, solutions are known that have no spatial symmetry,
namely the {\sl conformastatic class\/} [13]. The usual
physical interpretation of this class is a set of arbitrarily
placed extremal black holes that remains static because the
gravitational attraction is balanced by the electric repulsion between
the charges (which all have equal sign). But by appropriate choice of
integration constants one can eliminate the asymptotically flat region
and obtain a solution that is static and inhomogeneous in space, and that
can be thought of as a BR-type universe containing a number of extremally
charged black holes.
In its Riemannian analytic extension one can identify the static direction
with the BR $z$-direction, and consider the solution as inhomogeneous in
imaginary time. This will then give use the desired instanton.

The Lorentzian conformastatic metric and Maxwell field
have the form
$$ds^2_{CS} = V^2 d\sigma ^2 - V^{-2}{dt'}^2, \qquad
d\sigma ^2 = dX^2 + dY^2 + dZ^2 \eqno(6a)$$
$$ *F = \xi \wedge dV \eqno(6b)$$
where $V$ satisfies
$$ \nabla ^2 V = 0. \eqno(7)$$
Here the Laplacian $\nabla ^2$ is evaluated in the flat metric,
$d\sigma^2$, and $\xi$ is the Killing form corresponding to time translation
symmetry. To obtain a corresponding Riemannian solution we again
replace $t'$ by $it$ and write the solution suitable for
our purpose as
$$ds^2_{CS} = V^2 d\sigma ^2 + V^{-2}{dt}^2, \qquad *\!F = \xi \wedge dV
 \eqno(8a)$$
$$ V = \textstyle \sum \displaystyle M_i /\rho _i,
\qquad \rho _i = \sqrt {(X - X_i)^2 + (Y-Y_i)^2
+ (Z-Z_i)^2} \eqno(8b)$$
with $(X_i,Y_i,Z_i), \ i = 1...n$ denoting $n$ different points in $\R^3$.

Whenever $(X,Y,Z)$ approaches a particular, say the
$j$th, of the $n$ points, the $j$th term in the sum dominates, so that
$V$ approaches $M_j/\rho _j$. Thus the metric takes on
the BR form of Eq. (5) in the limit $\rho _j \rightarrow 0$. Similarly, in
the limit $(X,Y,Z) \rightarrow \infty$ we have $V \rightarrow
M_\infty/\rho$, where $\rho^2 = X^2 + Y^2 + Z^2$ and $M_\infty =\sum M_i$.
In this case the metric also
takes on the BR form (5), but in the limit $\rho \rightarrow \infty$.
Thus the geometry and field described by Eq. (8a) interpolate between the
$n+1$ BR universes corresponding to these asymptotic regions.
The flux parameter of each of the $n$ universes in the limit
$\rho _j \rightarrow 0$ is $M_j$,
and that of the $n+1$st universe, in the  limit $\rho \rightarrow \infty$,
is $\sum M_i$,
as expected from flux conservation.

The Maxwell tensor $F$ of this solution is not given directly by Eq. (8a),
but by way of its dual, $*F$. In the Lorentzian spacetime, replacing
$F$ by $*F$ (i.e., interchanging $E$ and $B$) yields another
Maxwell-Einstein solution with the same geometry. But only the solution
with a magnetic field
produces a real instanton, because the electric field becomes
imaginary under the replacement of $t'$ by $it$. (This is clear from
the transformation of the components $F^{0i}$, or from the fact that
the Killing form $\xi$ becomes imaginary under this replacement).
The magnetic field, on the other hand, remains real because  the
$\epsilon$-tensor, which defines the dual, also
picks up some imaginary components in the
analytic continuation. Alternatively, and more simply,
we can regard $\xi$ as the real Killing vector of the Riemannian
manifold and $*$ as its normal dual operation. Any electric fields
``induced" in the Riemannian development will then be
real\footnote{$^5$}{The situation is
analogous to particle motion in real vs. imaginary time. The particle
has a real momentum in the classically allowed region, and a real
momentum in the classically forbidden region after the potential has
been turned ``upside down"; but the two motions can connect only
at the turning point, where the momentum vanishes.}, and
satisfy the Riemannian Maxwell equations (whose main difference from
the Lorentzian Maxwell equations is that there is not the negative
sign usually associated with Lenz' law).

As usual with magnetic charges, the field $F$ of Eq. (8a) does not come
from a globally defined vector potential, but its dual can be so
derived (the usual restrictions on existence of potentials in wormhole
spacetimes [14] do not apply to these extremal configurations),
$$ *F = dC, \qquad C = V \xi . \eqno(9)$$

Finally we note that our instanton is geodesically
complete with finite Maxwell field everywhere, because the coordinates
of Eqs.~(8) are good everywhere except in the limits $\rho _j \rightarrow 0$
and $\rho \rightarrow \infty$, where
the geometry is asymptotically BR, that is $S^2 \times H^2$, and the
Maxwell field is finite.

\vskip 0.5truein
{\centerline{IV. THE ACTION}}
\vskip 0.25truein
\nobreak

The instanton of the last section has infinite extent in
the $t$-direction, and is homogeneous in that direction.
(Locally this corresponds to an axis of the asymptotic BR universes,
that is, a $z$-direction in the metrics of
Section II.)  The instanton can have a finite action only
if the contribution from any range of $t$ is independent of that
range --- that is, the contribution must have the form of a suitable
surface integral. It should also depend only on the number and
values of the flux parameters $M_j$, and not on the location of the
points $(X_j, Y_j, Z_j)$. This is so because our
instanton is an analytic continuation of an Einstein-Maxwell spacetime
that is static in the region considered (``static" here includes vanishing
of the electromagnetic momentum variable, {\it i.e.,} of the electric field).
The action of this
spacetime therefore consists of a potential energy term only. But for extremal
black holes, the gravitational
attraction is balanced by electromagnetic repulsion. Therefore this
potential, and hence the action, should not change with the location
of the holes.

The ``Euclidean" action, $\hat I$, for the Einstein-Maxwell fields is given by
[15]
$$-16\pi \hat I = \int (R - F_{\mu \nu} F^{\mu \nu})\sqrt{g}\ d^4x +
2\oint K \sqrt{^3g}\ d^3x +c.\eqno(10)$$
Here the surface integral over $K$ (the trace of the surface's
second fundamental form) is  needed to convert
the Einstein-Hilbert action, $\int R \sqrt{g}\ d^4x$, into the first-order
form appropriate for quantum gravity. The term $c$ is included to annul
the action of the initial state. (If the
initial state is flat space one usually puts $c = -2\oint K^0 \sqrt{^3g}\
d^3x$,
where $K^0$ is the value of $K$ when the boundary is embedded in flat space.)
The analogous normalization in the case of tunneling through a potential
fixes the energy of the classical initial state at zero, because only
then is the exponential fall-off of the WKB wavefunction given correctly
by $\exp(-\hat I/\hbar)$.

For an Einstein-Maxwell solution
the Einstein-Hilbert action $\int R \sqrt{g} d^4x$
vanishes, since the Maxwell stress-energy is traceless.  Similarly,
if $*F$ is derived from a potential, Eq. (9), then the
Maxwell part of the action
can be converted into a volume integral that vanishes when evaluated on
a solution, plus a surface integral :
$${1\over 2}\int F_{\mu \nu} F^{\mu \nu}\sqrt{g}\ d^4x = \int *F \wedge F =
\int dC \wedge F = \oint C\wedge F - \int C \wedge dF $$
Since $dF = 0$ by Maxwell's equations, the action becomes the pure
boundary integral
$$-8\pi \hat I = \oint (K + C_\mu * \! F^{\mu \nu} n_\nu)
\sqrt{^3g}\ d^3x + c.\eqno(11)$$

For the boundary three-surfaces we choose the cylinders with ``mantles"
$\rho = P$, $t~\in~(-T,T)$ and $\rho_j = P_j$, $ t \in (-T,T)$, and
``top" and ``bottom" surfaces $\rho \leq P$, $\rho_j \geq P_j$, $t~=~T$
resp.~$-T$, all in the limit
$P \rightarrow \infty$, $P_j \rightarrow 0$, $T \rightarrow \infty$.
The top and bottom are totally geodesic and normal to $C$ in the
gauge of Eq. (9). Hence both terms in the integrand of Eq. (11) vanish there.

On the mantle surfaces, $\rho = P$ resp. $\rho_j = P_j$,
we find, in the limit
$P \rightarrow \infty$ resp. $P_j \rightarrow 0$,
$$K = -1/M_j + O(P_j^2)\quad {\rm resp.}\quad 1/M_{\infty}+ O(P^{-2})
\eqno(12a)$$
and $$C_\mu *\! F^{\mu \nu} n_\nu =  1/M_j + O(P_j^2)\,
\quad {\rm resp.}\quad -1/M_{\infty}+ O(P^{-2}).\eqno(12b)$$
Since $K \neq 0$, neither of these limits represents totally
geodesic surfaces as would be appropriate for a bounce solution.
But this instanton does not involve a bounce: both ``initial" and
``final" states are reached only asymptotically in ``imaginary time."
Hence it is appropriate to evaluate the action in both regions
in the way one evaluates it in the
one asymptotically Euclidean region of bounce-type instantons.
That is, we take the limits and choose $c$ in such a way that
we obtain zero when the action is evaluated for the initial,
{\sl single} BR universe of Eq. (5), for which
$M = M_\infty = M_1$ and all other $M_j = 0$. We can achieve this
by taking the limits
$P \rightarrow \infty$, $P_j \rightarrow 0$ before the limit
$T \rightarrow \infty$, and setting $c = 0$. Because the terms
of Eqs. (12a) and (12b) then cancel for {\sl any\/} set of $M_j$'s,
the contribution to the action from the mantle surfaces always
vanishes as well.

There is, however, a nonzero contribution to the action from
the cylinder's two-dimen\-sional ``edges" at $\rho = P$, $\rho_j = P_j$,
$t = \pm T$. There the curvature $K$
has a $\delta$-function behavior, because the
boundary surface makes a sharp $90^{\circ}$ turn. The contribution
to the integral of Eq~(11) from the $j^{\rm th}$ pair of these edges
is $2 ({\pi \over 2})  A_j$, where $A_j$ is the area
of the edge $P_j \rightarrow 0$, that is $4\pi {M_j}^2$
as given by Eq.~(8). From
the sum of these contributions we must again subtract the value of a term
of this type for the
single BR universe. Thus the final result for the ``Euclidean" action is
$$\hat I = {1\over 8}({A_{\infty}} - \textstyle \sum \displaystyle {A_j})=
{\pi\over 2} [(\textstyle \sum \displaystyle  {M_j})^2
- \textstyle \sum \displaystyle {M_j}^2]. \eqno(13)$$
(Of course this does not take
into account any contribution to the action due to possible modifications
of (10) by topological invariants.)
This finite value of the action makes the solution (8) a proper
instanton.

It is curious that $\hat I$ is half the difference
in the entropies of the black holes whose asymptotic interiors are
approximated by these BR universes, so that the square of the
WKB wavefunction agrees with the probability $\exp(\Delta S)$ associated by
statistical mechanics with a fluctuation in which the entropy $S$
deviates by $\Delta S$ from its equilibrium value.\footnote{$^6$}{For pointing
this out I thank T. Jacobson and F. Dowker, who also suggested the
interpretation given to Eq. (13) in the discussion, below.}

The instanton and the corresponding WKB wavefunction describe the splitting
of one BR universe into two or several. An analogous instanton would be
expected to describe the splitting of an extreme Reissner-Nordstr\"om
wormhole into two or several, agreeing with the BR instanton in the
extreme, cylindrical interiors. The action of this instanton can have
a finite value, since the above shows that there is only a finite contribution
from the infinitely long cylinder. (In fact, it is likely that the action
in the wormhole case is similar to (13): although the corresponding
instanton is not known, to evaluate (11) one only needs the geometry
near its boundary surfaces, which one can estimate.)
\eject
\vskip 0.5truein
{\centerline{V. DISCUSSION}}
\vskip 0.25truein

The numerical value of the instanton action allows one to calculate
such quantities as the probability of wormhole (or BR universe)
splitting, or the relative probabilities of different numbers of wormholes
(or BR universes) being
present in the fluctuating state characterized by a given mass (or flux)
parameter. Eq.~(13) suggests that the process of one extremal wormhole
breaking into many is suppressed by $\exp (-2\hat I/\hbar)$, that is, according
to how badly the second law of black hole thermodynamics is broken. Thus
the present calculation supports the idea that the topology-changing
process is important on the Planck scale, for example when
$M_{\infty}$ is of the order of a Planck mass.

What other information  can we read out of the instanton?
To understand how a Lorentzian spacetime describes the development of
geometry and fields from an initial to a final state it is appropriate
to ``slice up" the spacetime by non-intersecting spacelike surfaces.
The same procedure is
appropriate for an instanton, except that the spacelike requirement is
no restriction, and that the surfaces typically cannot all be regular
if topology change is involved. The resulting ``imaginary time history"
shows how the fluctuation comes about in the sense that it provides a
sequence of geometries and fields that are most likely to be present
during the fluctuation.

For our instanton, a particularly simple slicing is by
surfaces $V =$ const. If we suppress the trivial axial ($t$-) direction,
the corresponding 2-surfaces are easily visualized as embedded in
three-dimensional space: except for the conformal factor $V^2$
they are simply the equipotentials of $n$
charges $M_j$. Thus, to visualize, say, the fluctuation history of one parent
BR universe into two equal daughters, think of the equipotentials of two
equal charges.
At large distances we have  spheres, corresponding to the parent
universe.  By applying the correction prescribed by
the conformal factor $V^2$ in Eqs.~(8) we find that their true radius
is $2M_1$.  As we move inward, the spheres distort and pinch off, forming
two surfaces that in the limit $\rho _1 \rightarrow 0$
or $\rho _2 \rightarrow 0$ again become spherical --- the daughter
universes, each of true radius $M_1$. This particular slicing is
distinguished by the vanishing of the electric field everywhere.
The magnetic field, being in the $t$-direction perpendicular to the
2-surfaces, could be visualized as a density of conserved points that avoid
the pinch-off region, where $dV = 0$ in Eq.~(6b).

The fields and geometries of this slicing may also be useful for
suggesting minisuperspaces in which such topology-changing processes
can be further investigated.

\vskip 0.5truein
{\centerline{Acknowledgments}}
\vskip 0.25truein
\nobreak
I am grateful to  E. Sch\"ucking and P. Hajicek for useful discussions
and references. I also thank T. Jacobson, R. Sorkin, F. Dowker, K. Pirk,
G. Gilbert and
J. Difelici for the interest they have taken in this work.
It was supported in part by National Science Foundation grant
No.~PHY-9020611-01.

\eject
\vskip 0.5truein
{\centerline{REFERENCES}}
\vskip 0.25truein

\frenchspacing
\item{[1]} J. A. Wheeler, {\it Phys. Rev.} {\bf 97}, 511 (1955) and
\item{ } {\it Geometrodynamics}, Academic Press,
New York, 1962, and in
\item{ } E. Nagel, P. Suppes and A. Tarski, {\it Logic, Methodology and}
{\it Philosophy of Science: Proceedings of the 1960 International Congress},
Stanford University Press 1962 p.~368~ff.

\item{[2]} R. Geroch, {\it J. Math. Phys.} {\bf 8}, 782 (1967).

\item{[3]} S. Coleman, {\it Phys. Rev. D} {\bf 15}, 2929 (1977);

C. Callen and S. Coleman, {\it Phys. Rev. D} {\bf 16}, 1762 (1977).

\item{[4]} R. Sorkin, {\it Intern. J. Theor. Phys.} {\bf 25}, 877 (1986).

\item{ }J. Friedman in {\it Conceptual Problems of Quantum Gravity},
A. Ashtekar and J. Stachel, eds., Birkh\"auser 1991.

\item{[5]} U. Gerlach, {\it Phys. Rev.} {\bf 177}, 1929 (1969).

\item{ }J. A. Wheeler in {\it Analytic Methods in Mathematical Physics},
R. D. Gilbert and R.~Newton, eds., Gordon and Breach 1970.

\item{[6]} D. Brill, {\it Foundations of Physics} {\bf 16}, 637 (1986).

\item{[7]} D. Garfinkle and A. Strominger, {\it Phys. Lett. B}
{\bf 256}, 146 (1991).

\item{[8]} J. B. Hartle and S. W. Hawking, {\it Phys. Rev. D}
{\bf 13}, 2188 (1976).

\item{[9]} B. Carter ``Black Hole Equilibrium States" in {\it Black Holes},
C. deWitt and B. S. deWitt, eds., Gordon and Breach 1973.

\item{[10]} A. Strominger, ``Baby Universes" in {\it Particles, Strings and
Supernovae}, A. Jevicki and C-I. Tan, eds., World Scientific 1989.

A. Auerbach, S. Kivelson and D. Nicole, {\it Phys. Rev. Let.}
{\bf 53}, 411 (1984).

\item{[11]} T. Levi-Civita, {\it R.C. Accad. Lincei} (5) {\bf 26}, 519 (1917).

B. Bertotti, {\it Phys. Rev.} {\bf 116}, 1331 (1959).

I. Robinson, {\it Bull. Akad. Polon.} {\bf 7}, 351 (1959).

\item{[12]} R. Lindquist, {\it The Two-Body Problem in Geometrodynamics},
doctoral dissertation, Princeton University 1959, p. 150.

\item{[13]} D. Kramer, H. Stephani, E. Herlt and M. MacCallum,
{\it Exact Solutions of Einstein's Field Equations},
Cambridge University Press 1980.

\item{[14]} Y. Choquet-Bruhat, C. DeWitt-Morette with M. Dillard-Bleick,
{\it Analysis, Mani\-folds and Physics} North-Holland 1982, p. 271.

\item{[15]} S. W. Hawking in {\it General Relativity, An Einstein Centenary
Survey}, Hawking and Israel, eds., Cambridge University Press 1979.

\end